\documentclass[twocolumn,showpacs,preprintnumbers,amsmath,amssymb]{revtex4}
\usepackage{graphicx}
\usepackage{dcolumn}
\usepackage{bm}

\newcommand{\BE}{\begin{equation}}
\newcommand{\EE}{\end{equation}}
\newcommand{\be}{\begin{equation}}
\newcommand{\ee}{\end{equation}}
\begin{document}

\preprint{to be submitted}

\title{Dynamics of Domain Growth in Self-Assembled Fluid Vesicles}
\author{Mohamed Laradji$^{1,3}$ and P. B. Sunil Kumar$^{2,3}$}
\affiliation {$^1$Department of Physics, The University of Memphis, 
Memphis, TN 38152 \\
$^2$Department of Physics, Indian Institute of Technology Madras,
Chennai 600036, India\\
$^3$MEMPHYS--Center for Biomembrane Physics, University of Southern Denmark, DK-5230, Denmark}

\date{\today}

\begin{abstract}
The dynamics of phase separation in multi-component bilayer fluid vesicles is investigated 
by means of large-scale dissipative particle dynamics. The model explicitly accounts for
solvent particles, thereby allowing for the very first numerical investigation
of the effects of hydrodynamics and area-to-volume constraints.
We observed regimes corresponding to coalescence of flat patches, budding and vesiculation, 
and coalescence of caps. We point out that the area-to-volume constraint has a strong influence on 
crossovers between these regimes. 
\end{abstract}

\maketitle

Apart from partitioning the
inner and outer environments of the cell, biomembranes also act as support for
complex and specialized molecular machinery, crucial for various
physiological functions and trans-membrane transport~\cite{alberts}.
It is believed that membranes  maintain in-plane compositional organization that is
essential for its function.
Though biological membranes are very complex and can in general be far
from equilibrium, knowledge of
the equilibrium properties of simple model membranes will be extremely useful
in providing understanding of this molecular machinery.
Such a study should also be essential for developing new applications
involving liposomes. This has recently resulted in a surge of theoretical and
experimental investigations on domain formation in multicomponent
vesicles~\cite{lipowsky92,lipowsky93,taniguchi96,kumar96,kumar98,kumar01,bagatolli01,veatch02,baumgart03}.

The dynamics of in-plane demixing of multicomponent vesicles 
into coexisting phases is richer than its counterpart in bulk systems.
There are several reasons for this: (i) the phase separation process 
is strongly coupled to the shape dynamics of the vesicle,
(ii) the viscosities of the lipid bilayer and that 
of the embedding solvent are different, and 
(iii) the membrane can be impermeable to 
solvent, resulting in a constraint on the vesicle area-to-volume ratio. 
The purpose of this letter is to investigate the effect of 
the above on the dynamics of multicomponent fluid vesicles.

Phase separation of vesicles following a quench to the two-phase
region has previously been considered by means of a generalized time-dependent Ginzburg-Landau model
~\cite{taniguchi96}, and dynamic triangulation Monte Carlo 
model~\cite{kumar96,kumar98,kumar01}. The later study showed a marked
departure of the phase separation in multicomponent
fluid vesicles from their Euclidean counterparts as a result of coupling between 
curvature and composition.  In particular, it was shown that at intermediate times,
the familiar labyrinth-like spinodal network 
breaks up into isolated domains~\cite{kumar96,kumar98}. In the case of a tensionless closed 
vesicle, at late time these domains reshape into buds connected to the parent
vesicle by very narrow necks~\cite{kumar01}. Further domain growth proceeds via 
Brownian motion of these buds and their coalescence. 

Recent advances in experimental techniques, like the two photon fluorescence 
and confocal microscopy,
has made it possible to study phase separation on 
fluid vesicles~\cite{bagatolli01,veatch02,baumgart03}.
These experiments reported 
structures with many domains 
which are more akin to caps than fully developed buds. 
A natural question then is: Does coarsening 
in multicomponent fluid vesicles proceed via a kinetic pathway similar
to that predicted by the recent simulations?

Previous studies of the dynamics of phase separation in fluid membranes
did not take into account the 
following important features of a real lipid vesicle: (i) the presence of 
solvent,(ii) the constraint of area-to-volume ratio and (ii) the freedom to 
vesiculate. 
We report here the first large scale simulation of phase separation dynamics
of a fluid vesicle model that accounts for these features.
This study, based on dissipative particle dynamics (DPD),
finds that the dynamics at all times is affected by the presence of
the solvent, and that the late time shape of the vesicle corresponds to that of a
surface decorated with caps.

Within the DPD approach~\cite{hoogerbrugge92}, the
vesicle is formed from the self-assembly of individual particles in an 
explicit solvent.  
The model parameters are chosen such that 
the membrane is impermeable to the solvent, 
thus allowing us to investigate the effect of conservation of both vesicle's 
area and encapsulated volume.  A lipid particle is modeled as a simple flexible 
amphiphilic chain built with four DPD particles; one hydrophilic
particle, simulating the lipid head group ($h$),
and three hydrophobic particles, simulating the lipid hydrophobic part ($t$).
The heads of the two types of lipids are denoted
by $h_A$ or $h_B$ and their tails are denoted by $t_A$ or $t_B$.
Water particles are denoted by $W$. We focus on the case where interactions are 
symmetric under exchange of $A$ and $B$ components, thus 
ensuring no explicit coupling between local composition and local curvature.
                                                                                
The position and velocity of each particle are denoted by
$({\bf r}_i,{\bf v}_i)$. Their time evolution is governed by Newton's
equations of motion ~\cite{gerhard}. There are three types of pairwise additive forces
acting on a particle $j$ by a particle $i$: (i) a conservative force, $F^{(C)}_{ij}$,
(ii) a dissipative force, $F^{(D)}_{ij}$ and (c) a random force, $F^{(R)}_{ij}$.
The conservative  force between two particles $i$ and $j$
is given by $F^{(C)}_{ij}= -a_{ij} \omega( r_{ij}) {\bf n}_{ij}$
where we choose $\omega(r)= 1- r/r_c $ for $r\le r_c$ and 
$\omega(r)=0$ for $r > r_c$ such that the forces are soft and
repulsive.  Here ${\bf r}_{ij}={\bf r}_j-{\bf r}_i$,
${\bf n}_{ij}={\bf r}_{ij}/|{\bf r}_{ij}|$ and  $r_c$ is the cutoff 
radius for the interaction. The hydrophobic and
hydrophilic interactions emerge from the relative interaction strengths $a_{ij}$.
With this, the parameters chosen for the simulation are
\be
a_{ij}=\frac{\epsilon}{r_c} \left(\begin{array}{cccccc}
   &h_A &t_A  &W  &h_B  &t_B\\
h_A&25  &200  &25 &100  &200\\
t_A&200 &25   &200&200  &100\\
W  &25  &200  &25 &25   &200\\
h_B&100 &200  &25 &25   &200\\
t_B&200 &100  &200&200  &25
\end{array}
\right),
\ee
where $\epsilon$ sets the energy scale.
The interaction parameters are chosen such that the amphiphiles self-assemble into bilayers 
and the two types of lipids are in the strong segregation regime.
A lipid particle integrity is ensured via a harmonic interaction between consecutive monomers
within a chain given by $ {\bf F}^{(S)}_{i,i+1}=-C( 1-r_{i,i+1}/b){\bf n}_{i,i+1}$,
where $C$ is a positive spring constant, and $b$ is a preferred bond length.
We use $b=0.45 r_c$ and $C=100\epsilon$.
The dissipative force originating from ``friction'' between particles is
given by,
${\bf F}^{(D)}_{ij}= -\gamma_{ij}\omega^2( r_{ij})({\bf n}_{ij}\cdot{\bf v}_{ij}){\bf n}_{ij}$ 
where $\gamma_{ij}$ is a friction parameter and ${\bf v}_{ij}={\bf v}_j-{\bf v}_i$.
 The random force is given by $
{\bf F}^{(R)}_{ij}= -\sigma_{ij} \omega( r_{ij})\zeta_{ij}(\Delta t)^{-1/2}{\bf n}_{ij}$,
where, $\Delta t$ is the time step in units of $\tau =(m r_c^2/\epsilon)^{1/2}$ with 
$m$ being the mass of a single DPD particle. All particles in our simulation have 
the same mass.  Here, $\zeta_{ij}$ is a symmetric random variable with zero mean 
and unit variance,
uncorrelated for different pairs of particles and at different time.
The dissipative and random forces are related to each other 
through the fluctuation dissipation theorem,
leading to the relation $ \gamma_{ij}=\sigma_{ij}^2/2k_{\rm B}T$. 
Unless otherwise specified, we used a fixed value $\sigma_{ij}=\sigma$ 
for all interacting pairs.
The pairwise nature of the 
dissipative and random forces ensures local momentum conservation leading to
correct long-range hydrodynamics~\cite{pagonabarraga98}.
                                                                                
\begin{figure}
\includegraphics[scale=0.35]{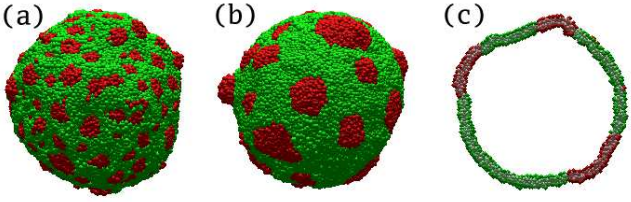}
\caption{Snapshots of a phase separating vesicle in case-I at (a) $t=100\,\tau$ and (b) $2000\,\tau$. (c) is a
slice taken at $5000\,\tau$.}
\label{caseIsnap}
\end{figure}

\begin{figure*}
\includegraphics[scale=0.45]{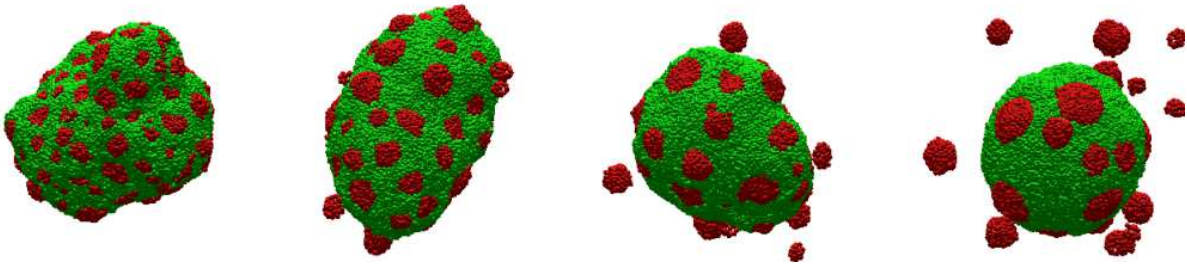}
\caption{Snapshots of a phase separating vesicle in case-II.
Snapshots from left to right are at $t=100\,\tau,\, 400\,\tau, \, 1000\,\tau$ and 
$4000\,\tau$, respectively. 
Analysis show that the detached objects are full vesicles.} \label{caseIIsnap}
\end{figure*}
                                                                                
In our simulation the following parameters were used
$\sigma=3.0(\epsilon^3 m/r_c^2)^{1/4},\,\Delta t=0.05 \tau$ and $k_B T = \epsilon $.
With the parameters chosen, the estimated bending modulus, $\kappa\sim 10 k_B T$, and
line tension, $\lambda \sim 10^{-17} J/\mu m$, are in reasonable agreement with
estimates for lipid membranes~\cite{lipowsky95}.
We use $16,000$ lipid chains in a simulation box of 
$80 \times  80  \times  80\, r_c^3$ 
with DPD-particle number density of $3 r_c^{-3}$. The
number of water particles inside the vesicle, when it is not deflated, is 
about $138,400$. The total number of DPD particles corresponds to 1,536,000~\cite{computer_time}.

\begin{figure}
\includegraphics[scale=0.7]{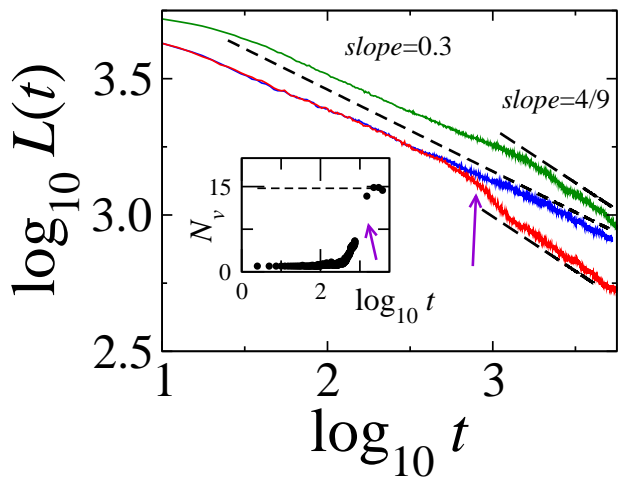}
\caption{Net interface length as a function of time in case-I (blue curve) and
case II (red curve). The green curve corresponds to case II with low line
tension (see text for details of corresponding interactions). 
The dashed lines and the dot-dashed line
have slopes of $0.3$ and $4/9$, respectively. 
In the inset, the number of vesiculated domains is shown as a function of 
time for case-II with high line tension (corresponding to the red curve).
The arrows point to the time regime during which budding and 
vesiculation occur in case II with high line tension.}
\label{interface}
\end{figure}

Bilayers and vesicles can be  
self-assembled in a DPD simulation~\cite{yamamoto02,yamamoto03}. 
In order to save computer time, 
we prepare our vesicles starting from an almost closed 
configuration, composed of a 
single lipid component, and allow it to equilibrate. 
This approach, allows for the equilibration of the lipid surface 
density within the two monolayers through the edge of the membrane.  
We find that the vesicle closes within  about 50,000 DPD steps. 
Within this model parameters, we found that the number of solvent particles inside
the vesicle and the number of lipid particles in the inner and outer
leaves of the vesicle remain constant.
This implies that the vesicle is impermeable to the solvent and flip-flop events are rare.
The vesicle prepared in this manner is found to be 
nearly spherical with about $140,000$ solvent particles inside it.
We refer to configurations with this inner volume as case-I.
A vesicle with excess area can then be prepared by transferring a certain amount of solvent
particles from its inner to outer regions. We refer to as case-II
the equilibrated configurations obtained after $20\%$ of the inner solvent particles are transfered. 
We found that in case-II, as well as in case-I, the solvent density is the same inside and outside the vesicle.

To study the phase separation, we randomly label each lipid particle in 
the bilayer as type $A$ or $B$, such that the relative composition in 
each layer is the same.  
Coarsening  is subsequently monitored by computing the distribution of
cluster sizes of the minority component and the total interface 
length between the domains. 
We will focus on the case of off-critical quenches with
the volume fraction of the B-component equal to 0.3.

\begin{figure}
\includegraphics[scale=0.5]{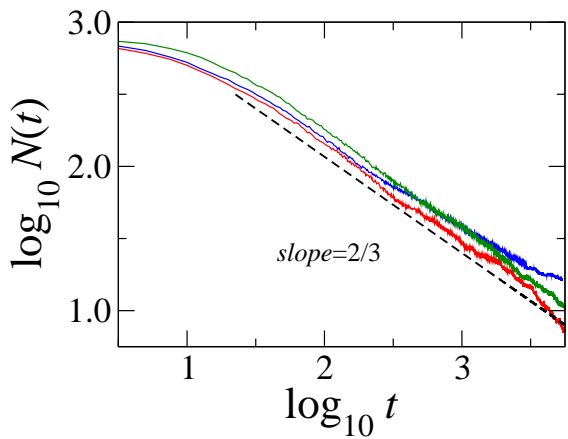}
\caption{Number of clusters on the vesicle as a function of time. Curve colors 
correspond to those in Fig.~\ref{interface}. The slope of the dashed line 
is 2/3.} 
\label{cluster}
\end{figure}

Snapshots of the time evolution of two component vesicles in 
cases I and II are shown in Figs.~\ref{caseIsnap} and \ref{caseIIsnap},
respectively. As shown in Fig.~\ref{caseIsnap} (c), domains on either side of the membrane
are in register. This inter-layer alignment since early stages of the phase separation 
dynamics.
The net interface length is shown in Fig.~\ref{interface}.
During relatively early times, in both cases I and II, domains are 
flat circular patches.  The net interface length, $L(t)$, during this regime is 
independent of the area-to-volume ratio, and has the form $L(t)\sim t^{-\alpha}$, with 
$\alpha\approx 0.3$, as shown in Fig.~\ref{interface}.
In this regime, the number of clusters on the vesicle, $N_c$, 
is also found to obey a power law, $N_c\sim t^{-\beta}$, with $\beta \approx 2/3$. 
This is depicted in Fig.~\ref{cluster}. 

Since the amount of $A$ and $B$ components on the membrane are conserved,  
we have $N_c(t)\sim {\cal A}_B/R^2(t)$, where $R$ is the 
average domain size and ${\cal A}_B$ is the total area occupied by the $B$ patches. 
Since the patches are flat $R(t)\sim {\cal A}_B/L(t)$. 
Thus one expects $\beta=2\alpha$. The data shown in Figs. 
\ref{interface} and \ref{cluster} are consistent with this relation.
If the diffusion coefficient of disks moving on the membrane depends 
on the domain size as $D_d\sim 1/R(t)$
then coalescence of domains 
leads to $\alpha=1/3$ and $\beta=2/3$~\cite{furukawa85}. These exponents can also be the 
result of evaporation-condensation mechanism, as predicted by 
Lifshitz and Slyozov~\cite{lifshitz62}. 
However, we could monitor many events of patch coalescence in our simulation.
We thus believe that the mechanism of domain growth during this regime 
is coalescence of patches~\cite{comment}. 

After about $t=400\tau$, the dynamics in cases I and II clearly departs from 
each other, as shown in Fig.~\ref{interface}.
In case-I, although there are some caps formed, 
coarsening proceeds mainly through coalescence of 
flat circular patches (see Fig.~\ref{caseIsnap}). 
Further capping is suppressed by the lateral tension 
on the membrane due to volume constraint. Similar capping-induced tension
was also seen in earlier Monte Carlo simulations~\cite{kumar98}.
In case-II, where there is more excess area, domains reshape into caps.
The presence of {\em excess area} allows  a fraction of the caps to
further reshape into buds which then vesiculate. 
The buds vesiculate during a short time implying that the energy barriers 
involved are very small.  We confirm this by performing simulations on a 
vesicle with large excess area having a single $B$ component domain occupying 
$12\%$ of the total area. Once the bud is formed, it is found to vesiculate within
a time period of $10 \tau$.  This budding and vesiculation results in a 
marked decrease in the net interface length $L(t)$ over a short period of time
as shown in Fig.~\ref{interface}. Pinching mechanism of vesicles
was recently studied in detail using DPD simulations~\cite{yamamoto03}.

All detached vesicles are composed of the $B$ component. Budding occurs 
towards both inner and outer regions of the vesicles,
although most buds are found in the outer region.  Inward 
budding is due to the bilayer nature of the vesicles and cannot be observed 
if the vesicle is modeled by a single surface. 
In the absence of explicit coupling between mean curvature and composition,
inward vesiculation of a $B$-domain results when the area of the domain
on the inner monolayer exceeds its area on the outer monolayer.

Vesiculation results in a reduction of area-to-volume ratio
of the parent vesicle leading it to acquire a more spherical shape. 
The lateral tension resulting from this prevents further
increase in curvature of caps.  
Coarsening now proceeds via coalescence 
of these caps. Since patch and cap coalescence, in a viscous medium, lead to
the same exponent, $\beta$, we do not see any marked change in the behavior of the curve
shown in Fig.~\ref{cluster} for case-II.
During this regime, the net interface length $L\sim t^{0.4}$, as 
shown in Fig.~\ref{interface}.
The exponent $\alpha\approx 0.4$ is between that due 
to coalescence of flat patches, i.e. $\alpha=1/3$, 
and that due to coalescence of buds having an interface length independent 
of the bud size, i.e. $\alpha=2/3$~\cite{kumar01}.
As can be seen in Fig.~2, domains remaining on the vesicle typically have a shape
closer to a hemispherical cap than a complete bud. 
In this case, the average interface length of a cap,
$l_c \sim (\kappa a/\lambda)^{1/3}$, where $a$ is the cap area. 
In the absence of further vesiculation, $N_c\, a$ is a constant of time.
We thus obtain for the net interface length $L\sim N_c l_c\sim t^{-4/9}$, 
consistent with our numerical results shown in Fig.~\ref{interface}.

\begin{figure}
\includegraphics[scale=0.35]{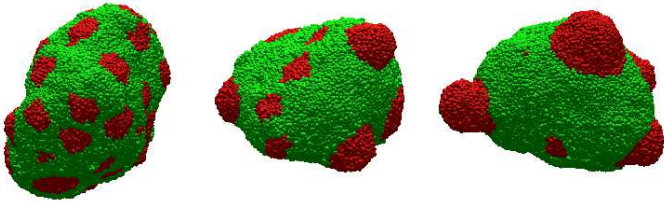}
\caption{Snapshots sequence for case-II, with line tension lower that in 
Fig.\ref{caseIIsnap}. Snapshots from
left to right correspond to $t= 1200$, $3700$, and $7500\ \tau$, 
respectively.} 
\label{lowtensIIsnaps}
\end{figure}

Now, let us discuss the effect of line tension on domain growth in case-II.
For a given $B$-cap of area $a$, surrounded by the $A$-phase, 
the bending energy of the cap scales with its radius of curvature, $r$, as $\kappa a/r^2$.
Since the perimeter of the domain decreases with decreasing $r$, the cap adopts a shape with
finite curvature. The domain area beyond which capping 
occurs is around $4\pi\kappa^2/\lambda^2$.
Let $t_1$ be the time required to reach this capping regime. 
Beyond $t_1$, domain growth proceeds via
cap coalescence.  In the presence of enough excess area, as in case-II, this 
coalescence may lead to caps with interface length comparable to the 
thickness of the bilayer. Vesiculation may then proceed~\cite{yamamoto03}.

For a $\lambda$, and $\kappa$, there is 
an area of the cap around $16\pi\kappa^2/\lambda^2$ 
beyond which a complete bud is formed, provided that there is enough excess area. 
Let $t_2$ be the average time necessary to reach this area. 
During the period between $t_1$
and $t_2$, growth should be mediated by coalescence of caps.
In case-II, discussed above, $t_2-t_1$ is too short to numerically detect this regime. 
We therefore performed
another set of simulations with a reduced value of the line tension, achieved
by choosing the interaction parameter $a_{ij}$ for $t_At_B$ and $h_Ah_B$ pairs 
to be $50\epsilon/r_c$,
Corresponding snapshots are shown in Fig.~\ref{lowtensIIsnaps}. As can be seen 
from this figure, vesiculation does not occur even at very late times. However, well defined
caps with finite curvature are prominent. 
The net interface length shows a clear crossover between the
patch coalescence regime and caps coalescence regime, as shown in Fig.~\ref{interface}.

In summary, we presented here the very first study of phase separation dynamics of 
self-assembled bilayer fluid vesicles with hydrodynamic effects.
We found rich dynamics  with crossovers that depend strongly on area-to-volume ratio and
line tension between the two coexisting phases. 
In particular, we found that vesiculation, when it happens, occurs during a short period of time. 
Another feature that was not possible
to account for in earlier simulations but has important consequences is the volume constraint. 
As a result of tension induced by this constraint,
the main vesicle eventually acquires a spherical shape decorated with caps, irrespective of initial 
area-to-volume ratio, a result seemingly born out in recent 
experiments~\cite{bagatolli01,veatch02,baumgart03}

The authors would like to thank O.G. Mouritsen and L. Bagatolli 
for useful discussions. We thank M. Rao and G.I. Menon for critical comments.
MEMPHYS is supported by the Danish National Research Foundation. Part of
the simulations were carried out at the Danish Center for Scientific Computing.


\begin{thebibliography}{00}
\bibitem{computer_time} A run up to 100,000 DPD time steps takes  330 cpu hours on
a 3.0 GHz Pentium IV. Data shown are averages over six independent runs for each case.
\end{thebibliography}
\end{document}